\documentclass[nofootinbib]{appolb}
\usepackage[top=20mm, bottom=30mm, left=15mm, right=15mm]{geometry}
\usepackage[usenames,dvipsnames]{color}

\newcommand{\beq}{\begin{equation}}     \newcommand{\eeq}{\end{equation}}
\newcommand{\beqa}{\begin{eqnarray}}    \newcommand{\eeqa}{\end{eqnarray}}
\newcommand{\bde}{\begin{description}}  \newcommand{\ede}{\end{description}}
\newcommand{\ben}{\begin{enumerate}}    \newcommand{\een}{\end{enumerate}}

\newcommand{\la}{\langle}               \newcommand{\ra}{\rangle}

\newcommand{\kT}{{k_{\rm B}T} }

\newcommand{\eqn}[1]{\beq{ #1 }\eeq}

\newcommand{\inv}[1]{{\frac{1}{#1}}}

\newcommand{\inRbracket}[1]{{\left({#1}\right)}}

\usepackage{graphicx, rotating}
\usepackage{epsfig}
\usepackage[abs]{overpic}
\usepackage{psfrag}
\usepackage{url}
\usepackage{amsmath,amssymb,xcolor}
\begin{document}
\title{Progressive quenching --- Ising chain models%
\thanks{Invited talk presented at the 30th Marian Smoluchowski Symposium (September, 2017, Krakow)}%
}
%
%
\author{Michael Etienne$^1$, Ken Sekimoto$^{2,1}$
\address{$^1$Gulliver, CNRS-UMR7083, ESPCI, 75231 Paris, France }
\\
{} 
\address{$^2$Mati\`{e}res et Syst\`{e}mes Complexes, CNRS-UMR7057, Universit\'e   Paris-Diderot, 75205 Paris, France}
}
\maketitle
\begin{abstract}
Of the Ising spin chain with the nearest neighbor or up to the second nearest neighbor interactions, 
we fixed progressively either a single spin or a pair of neighboring spins at the value they took.
Before the subsequent fixation, the unquenched part of the system is equilibrated. We found that, in all four combinations of the cases, the ensemble of quenched spin configurations is the equilibrium ensemble.
\end{abstract}
  
\section{Introduction}
In a system of many degrees of freedom we think of fixing/quenching suddenly the degree of freedom of one group after another. The value taken at the moment of quenching are registered as they are. The unquenched part either continues the prescribed dynamics or simply in contact with a heat bath.  The main interest is the statistics of the final quenched states. We note that this is not so-to-say glassy state because we fix completely the freedom.

For example, we extrude a molten liquid, such as of iron or polymer, then  
the extruded part  is suddenly quenched. And we are interested in the structure of the surface or inside. Another example may be the process of decision making in a community before a referendum. Some of the community members will make up their mind very early, and they influence more or less the other's opinion. Then the people progressively make up mind, by the time of vote.

Or, when we bake a pancake using a hot plate, the liquid in contact with the hot plate is first baked, then the solid part progressively grows upward. We are interested in the coarsening of the bubbles in the pancake. Perhaps the formation of human personalities may be another example. Upon the birth we have lot of flexibility. In the course of life we progressively fix our viewpoint or the way of thinking, and finally we can have a lot of prejudices.

We will call these type of processes the {\it Progressive Quenching} or {\it {\sf Pq}}, for short.
The fixed part acts as a boundary condition or as an external field on the unquenched part of the system. 
And when we suddenly fix some part of unquenched degrees of freedom, the boundary condition is updated.
As we wrote above, the quantity of interest in common is the final state and its statistics, when the whole system has been fixed.

Some time ago one of the authors studied this type of problem for diffusive Goldstone modes
\cite{phason-freezing-PhA}. 
In the context of quasicrystal, the so-called phason field obeys  
essentially the diffusion equation with non-conservative thermal noise.
And this field is quenched progressively from the left to the right with a fixed velocity, $V.$  
They found that the spatial spectrum {\it in the quenched sample} is  
modified from the equilibrium one over the length scale inferior to the diffusion length $\ell$, defined by the ratio, $D/V,$ where $D$ is the diffusion constant. In a simplified version in one-dimensional space with scalar phason field $\phi(x)$, this modification corresponds to the change of mean-square displacement (MSD), $\la |\phi(x)-\phi(x')|^2\ra$ from $\sim |x-x'|$ in equilibrium to $\sim |x-x'|^{3/2}/\ell^{1/2}$ for $|x-x'|\lesssim \ell.$ For $V>0$ the MSD is diminished by factor $(|x-x'|/\ell)^{1/2}$ due to the temporarily fixed boundary value of $\phi$ at $x=Vt$ which breaks the symmetry of this Goldstone mode. From the viewpoint of the stochastic process, this boundary condition renders the field $\phi(x,t)$ {\it in the unquenched part} to be martingale \cite{martingale-book} so that 
$E[ \phi(x,t)| {\{\phi(u):u\le Vt\}}]=\phi(Vt,t)$ holds approximately for $0<(x-Vt)\lesssim \ell,$ where $E[X|Y]$ is the expectation value of $X$ under fixed $Y.$

Very recently, one of the authors studied the {\sf Pq} of the globally coupled spin model
\cite{PQ-global-2017axv}. They observed the emergence of a martingale property.

All the above studies are either asymptotic or numerical. We here report our study of several exactly solvable models.
We ask if any martingale aspect appears in some observables.

\section{Model and protocol}
\subsection{Setup}
\begin{figure}[h]
\centerline{%
{\includegraphics[width=18.cm,angle=0.]{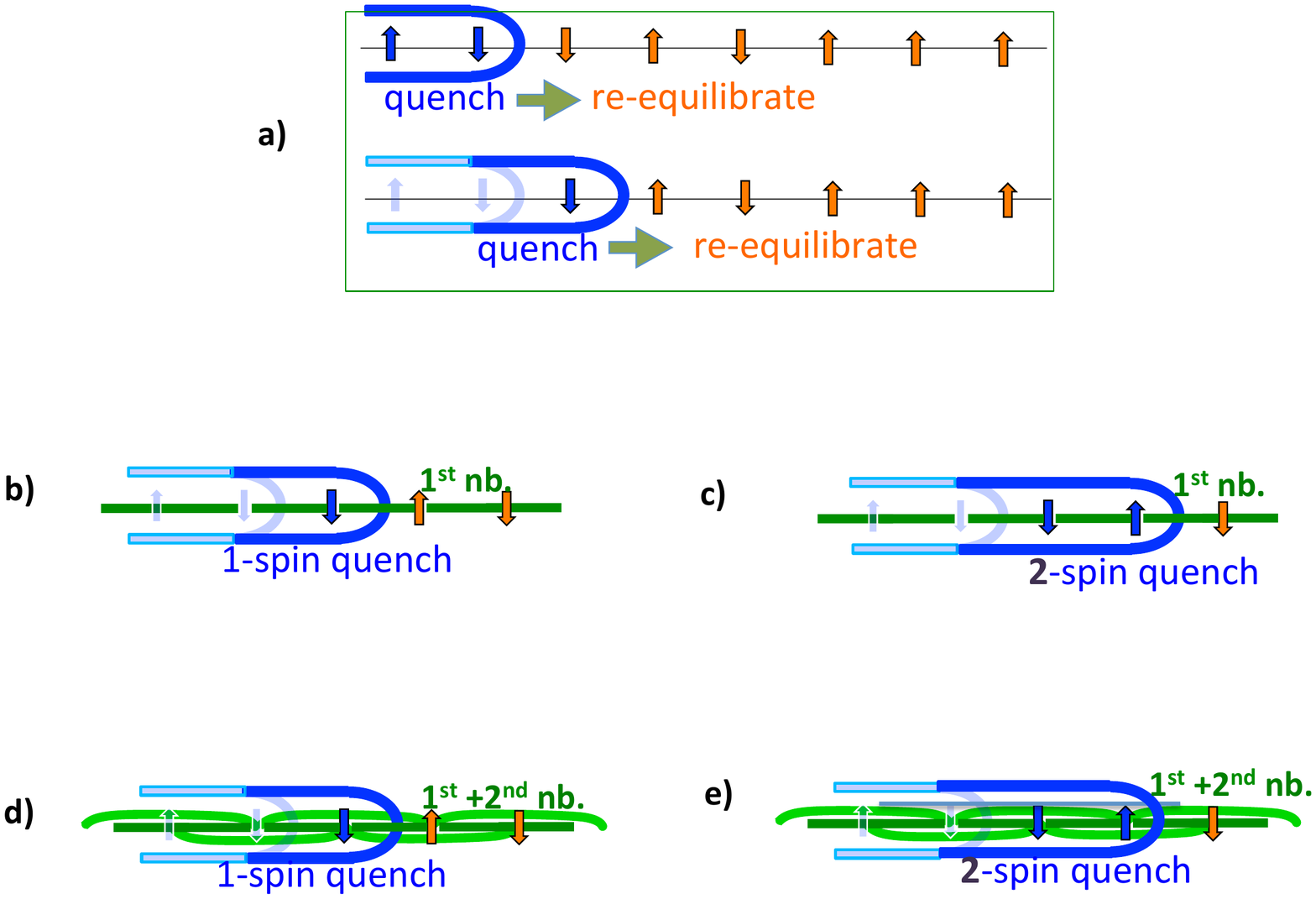}}  
}
\caption{a) Elementary iterative step of progressive quenching.  After the unquenched part is re-equilibrated a specified number of spins are fixed at their values when they took at the moment 
From b) to e) present different systems and different quenching units  In b) and c) the spins interact with their own first nearest neighbors, while in d) and e) the spins interact also with their second nearest neighbors 
In b) and d) a single spin is quenched at a time, while in c) and e) a pair of spins are quenched at a time }
\label{fig:schema}
\end{figure}

Fig.\ref{fig:schema} (a) presents the basic setup of the {\sf Pq} that we study in the present paper. 
We consider a chain of classical Ising spins.
After an event of quenching (see below) is done, the unquenched part is re-equilibrated.
Then a specified number of spins (a single spin in the case of Fig.\ref{fig:schema}(a)) are fixed at their values when they took at the moment.
This is the quenching event.  The values of spins fixed are, therefore, chosen from the equilibrium ensemble of the unquenched spins' configurations. Those spins are subject to the interactions with the {\it quenched spins} in addition to the interaction among the unquenched part.
We should note that this process is {\it not} quasi-static even though we completely re-equilibrate every time after quenching some spins. It is because the fixing of some spins implies to raise the barrier for the flipping of these spins 
so that the mean flipping interval exceeds the time-scale of observation/operation (see Chap.7.1 of \cite{LNP}).

We will study two Ising models. The one has the nearest neighbor interaction (Fig.\ref{fig:schema}(b) and (c)) and the other has the nearest and second-nearest interactions (Fig.\ref{fig:schema}(d) and (e)).
For the Ising chain with up to the second-nearest neighbor interaction, 
the energy $H$ can be written as 
\beq
- H= J_0\sum_{i=1}^{N-1} s_i s_{i+1}+J_1 \sum_{i=1}^{N-2} s_i s_{i+2}+h\sum_{i=1}^{N} s_i.
\eeq
If $J_1=0$ the system has only the nearest neighbor interaction.
For $J_1\neq 0$ the model including the second-nearest interaction can be mapped into the chain of spin-pair where the spin-pair has only nearest neighbor interaction. We introduce the composite variable, $\xi_p\equiv \{s_{2p-1},s_{2p}\}$ and regroup the energy $H$ for $N=2P$ as follows:
\beqa
\label{eq:pairing}
-H &=& \sum_{p=1}^{P-1}(J_0 [s_{2p-1} s_{2p}+s_{2p+1} s_{2p+2}]+J_1 [s_{2p-1} s_{2p+1}+s_{2p} s_{2p+2}]
+h [s_{2p-1}+s_{2p}])
\cr &\equiv & -\sum_{p=1}^{P-1} \mathcal{E}(\xi_p,\xi_{p+1}).
\eeqa
Then the second line on the r.h.s. is the nearest neighbor interaction between $\xi_p$ and $\xi_{p+1}.$
Though such pairing introduces apparent breaking of the system's translational symmetry by one spin, the system's physical behavior is intact.

We study two protocols of {\sf Pq}. The one quenches a single spin at one time (Fig.\ref{fig:schema}(b) and (d)) and the other quenches a pair of spins at one time (Fig.\ref{fig:schema}(c) and (e)).


\subsection{Transfer matrix and Markovian process along chain}
{In equilibrium the models of Ising spin chain are analytically treatable by the method of transfer matrix, as is described in the standard textbooks of statistical mechanics. The transfer matrix description allows to represent the canonical partition function as the discrete-time path integral over Markovian processes, where the time is the position along the chain. When the model has second-nearest neighbor interaction, the time is associated to each spin-pair. In Fig.\ref{fig:schema}, we see some similarity between the cases b) and e) because these system and protocol concern the single transfer matrix, of a single spin for b) and a pair of spin for e). }
Although we will not use the concrete expressions of the transfer matrix and its spectra, we will 
recall them to view its Markovian aspects.

For the Ising chain with the first-nearest neighbor interaction, the canonical partition function
$Z_{N}$ for the spins $\{s_1,\ldots,s_N\}$ reads (hereafter we use the energy unit so that $\beta=(\kT)^{-1}=1$)
\beq
Z_N=(1,1)\inRbracket{
   \begin{array}{cc}     M_{1,1} & M_{1,-1} \\      M_{-1,1} & M_{-1,-1} \\   \end{array}
}^{N-1}   \inRbracket{ \begin{array}{c}   e^{h} \\ e^{-h} \\   \end{array}  },
\eeq
where $M_{s,s'}=e^{J_0 s s' +h s}$ with the coupling constant $J_0$ and the external field, $h$.
When we are interested in the equilibrium probability,  ${\rm Prob}(s_j=s, s_k=s')$ with $1<j<k<N$,
we calculate 
\beq \label{eq:Pss}
{\rm Prob}_{\rm eq}
(s_j=s, s_k=s')=\inv{Z_N}
(1,1) M^{j-1} \mathcal{P}_s M^{k-j} \mathcal{P}_{s'}M^{N-k} 
\inRbracket{\begin{array}{c}e^{h}\\e^{-h}\\ \end{array}},
\eeq
where $\mathcal{P}_s$ is the projector matrix defined as 
$\mathcal{P}_1={\tiny 
\inRbracket{   \begin{array}{cc}  1 & 0 \\     0 & 0 \\   \end{array}}}$
and 
$\mathcal{P}_{-1}={\tiny \inRbracket{   \begin{array}{cc}  0 & 0 \\     0 & 1 \\   \end{array}}}$.
In practice we would take the limit of $k\to\infty$ and $N-k\to \infty$ in keeping the value of $k-j.$
Then the power $M^K$ with $K\to \infty$ can be replaced by $|0\ra {\lambda_0}^K\la 0 |$ with $\lambda_0$ being 
the largest eigenvalue of $M$ and $|0\ra$ and $\la 0 |$ are, respectively, the associated right and left eigenvectors.
The formula (\ref{eq:Pss}) allows to have the single-spin probability for $s$ if we sum over $s'$, and then allows to calculate the conditional probabilities as we wish, such as ${\rm Prob}_{\rm eq}(s_k=s' |s_j=s)
={\rm Prob}_{\rm eq}(s_j=s, s_k=s')/{\rm Prob}_{\rm eq}(s_j=s).$

In order to use the transfer matrix formalism in the model having second-nearest neighbor interactions, we introduce the four-space as 
$ \{\xi \}\equiv  \{(1,1),(1,-1),(-1,1),(-1,-1)\}.$ The partition function $Z$ is then given as 
$Z=\la\,| M^{P-1}|\, \ra$ with $\la\, |=(1,1,1,1)$ and $|\, \ra^t=(e^{J_0+2h},e^{-J_0},e^{-J_0},e^{J_0-2h}),$
and the components of the $4\times 4$ matrix $M$ are defined by $M_{\xi,\xi'}=e^{-\mathcal{E}(\xi,\xi')},$ 
where $\mathcal{E}(\xi,\xi')$ has been defined in (\ref{eq:pairing}).
By the same token as the nearest neighbor interacting chain, we can calculate any correlation function about $\xi$'s using this representation.

\section{Quenched ensembles of spin configuration}
Below we study the quenched ensemble of the spin chain for each case of Fig.\ref{fig:schema}(b)-(e). 
The section is not arranged in this order, rather in the order of increasing complexity of the argument.
Somehow surprisingly the conclusion is unique: all the quenched ensemble is the same as the equilibrium one for the given model, despite the non-equilibrium quenching operation.
Hereafter, we shall use the abbreviation, $P_{\rm eq}(s)$ for ${\rm Prob}_{\rm eq}(s)$  and 
$P_{\rm qu}(\xi,\xi')$ for ${\rm Prob}_{\rm qu}(\xi,\xi')$ etc.

\subsection{Ising chain with the nearest neighbor interaction quenched one-after-one spin (Fig.\ref{fig:schema}(b))
\label{subsec:PQ11}}
First of all we notice that the {\sf Pq} in this model is a Markovian process:
Suppose that  those spins $\{s_i\}$ with $\forall i\le i_0$ are already quenched. 
Hereafter, we assume that the total number of the spins $N$ is large enough that the effect of the both ends are negligible as long as the temperature is finite.
The equilibrium statistics of the unquenched spins $\{s_i\}$ with $i>i_0$ is influenced only by the state of the spin $s_{i_0}$. 
In the next step of quenching the state of $s_{i_0+1}$ to be quenched is given by the equilibrium conditional probability, $P_{\rm eq}(s_{i_0+1}|s_{i_0})$, which we can calculate using the transfer matrix technique.
Therefore, the conditional probability for the quenched spin configuration, $P_{\rm qu}(s_{i_0+1}|s_{i_0}),$
is given by,
\beq \label{eq:Pcond11}
 P_{\rm qu}(s_{i_0+1}|s_{i_0})= P_{\rm eq}(s_{i_0+1}|s_{i_0}).
\eeq
This is all what defines the statistics of the quenched sequence of spins.

Once we know the "transition probability" $P_{\rm qu}(s_{i_0+1}|s_{i_0}),$
we can find the single spin probability in the quenched sequence, $P_{\rm qu}(s_i).$ 
We admit that, after all the spins are quenched, the ensemble of the spin configurations is expected to have a translational invariance. Then $P_{\rm qu}(s)$ should satisfy a form of the Fredholm equation,
\beq \label{eq:fr11}
P_{\rm qu}(s)=\sum_{s_0}P_{\rm qu}(s|s_0)P_{\rm qu}(s_0),
\eeq
together with the normalization, $\sum_s P_{\rm qu}(s)=1.$ This equation  is the eigenvalue equation for the $2\times 2$ matrix, $P_{\rm qu}(s|s_{i_0}),$ with the eigenvalue of 1. By the way  with  (\ref{eq:Pcond11}) $P_{\rm qu}(s)=P_{\rm eq}(s)$ satisfies (\ref{eq:fr11}).
If we admit the uniqueness of the (normalized) solution for (\ref{eq:fr11}), we have\footnote{The other eigenvector of $P_{\rm eq}(s)$ is $\propto (1,-1)$ with the eigenvalue less than 1.}
\beq \label{eq:Ps11}
P_{\rm qu}(s)=P_{\rm eq}(s)
\eeq
Therefore, we arrive at the conclusion: the ensemble of the quenched spins are identical to the equilibrium one.

\subsection{Ising chain with the nearest neighbor interaction quenched two by two spins (Fig.\ref{fig:schema}(c))}
We will use the indexation of spin, $\ldots, s_1,s_2,s_3,s_4,s_5, \ldots = \ldots \xi_0, \xi,\ldots,$ where  $\xi_0=\{s_1,s_2\}$ and $\xi=\{s_3,s_4\}.$
Let us suppose that the quenching is operated on the spin pair, $(s_{2p+1}, s_{2p+2}),$ given the frozen configuration up to 
$s_{2p}$. 
We then have 
\eqn{\label{eq:14}
P_{\rm qu}(s_{3},s_{4}|s_{1},s_{2})=P_{\rm eq}(s_{3},s_{4}|s_{1}, s_{2}),}
which is analogous to (\ref{eq:Pcond11}). 
Because the equilibrium spins $(s_{3},s_{4})$ don't see $s_{1}$ if
the value of $s_{2}$, the r.h.s. is equal to $P_{\rm eq}(s_{3},s_{4}|s_{2}).$
It in turn means $P_{\rm qu}(s_{3},s_{4}|s_{1},s_{2})=P_{\rm qu}(s_{3},s_{4}|s_{2})=P_{\rm eq}(s_{3},s_{4}|s_{2}).$ By summing over $s_{4}$ we have
\eqn{P_{\rm qu}(s_{3}|s_{1},s_{2})=P_{\rm qu}(s_{3}|s_{2})=P_{\rm eq}(s_{3}|s_{2})}
Below we are going to show that the same form of relation holds for the shifted spin labels.
 \eqn{P_{\rm qu}(s_{4}|s_{2},s_{3})=P_{\rm eq}(s_{4}|s_{3})} 
{\it Derivation ---}\, 
We prepare the equality,
\beq
P_{\rm qu}(s_3,s_4|s_2)=P_{\rm eq}(s_3,s_4|s_2) = P_{\rm eq}(s_4|s_3) P_{\rm eq}(s_3|s_2) 
\eeq
Dividing each ends by the each side of 
 $P_{\rm qu}(s_{3}|s_{2})=P_{\rm eq}(s_{3}|s_{2}),$ which we mentioned above, we have
\beq
\frac{P_{\rm qu}(s_3,s_4|s_2)}{P_{\rm qu}(s_3|s_2)}=P_{\rm eq}(s_4|s_3)
\eeq
As the l.h.s. is identical to $P_{\rm qu}(s_4|s_2,s_3)$ we have 
\beq 
P_{\rm qu}(s_4|s_2,s_3)=P_{\rm qu}(s_4|s_3)=P_{\rm eq}(s_4|s_3) \,\,\,\,\, {\rm Q.E.D.}
\eeq
Once we have the "transition rates," $P_{\rm qu}(s_3|s_2)$ and $P_{\rm qu}(s_4|s_3),$
the stationary probabilities, $P_{\rm qu}(s_3)$ and $P_{\rm qu}(s_4)$ should satisfy
\beqa
P_{\rm qu}(s_3) &=&\sum_{s_2} P_{\rm eq}(s_3|s_2) P_{\rm qu}(s_2)
\cr
P_{\rm qu}(s_4) &=&\sum_{s_3} P_{\rm eq}(s_4|s_3) P_{\rm qu}(s_3).
\eeqa
This is a coupled Fredholm equation for the four-vector composited by the 
two-vectors, $P_{\rm qu}(s_2)$ (even-labeled sites) and $P_{\rm qu}(s_3)$ (odd-labeled sites).
If we admit the uniqueness of the normalized solution each for even and odd two-vectors\footnote{
It reduces to the eigenvalue problem, $P_{\rm qu}(s_4)=
\sum_{s_2}
(\sum_{s_3}P_{\rm eq}(s_4|s_3)P_{\rm eq}(s_2|s_2))
P_{\rm qu}(s_2),$ for the same two-vector, $P_{\rm qu}.$
The second eigenvector of the matrix, $(\sum_{s_3}P_{\rm eq}(s_4|s_3)P_{\rm eq}(s_2|s_2)),$ ---  the one other than the equilibrium one --- is $\propto{(1,-1)}$ with the eigenvalue less than 1.},
we conclude
\beq
P_{\rm qu}(s_3=\sigma) =P_{\rm qu}(s_4=\sigma) =P_{\rm eq}(\sigma).
\eeq
Therefore, the quenched ensemble is the same as the equilibrium ensemble 
in spite of the operation of {\sf Pq} that apparently break the translational symmetry.

\subsection{Ising chain with up to the second-nearest neighbor interaction quenched one-after-one spin (Fig.\ref{fig:schema}(d))}
For the purpose of the simplicity of notation, we again introduce the indexation of spin, $\ldots, s_1,s_2,s_3,s_4,s_5, $  $\ldots $  $= \ldots \xi_0, \xi,\ldots,$ where  $\xi_0=\{s_1,s_2\}$ and $\xi=\{s_3,s_4\}.$ (Because of the translational symmetry under the shift by a single spin,
we could also assign like $\xi_0=\{s_2,s_3\}$ and $\xi=\{s_4,s_5\}.$ )
The protocol of {\sf Pq} means $P_{\rm qu}(s_3|s_1,s_2)=P_{\rm eq}(s_3|s_1,s_2)$ and $P_{\rm qu}(s_4|s_2,s_3)=P_{\rm eq}(s_4|s_2,s_3).$ 
We will study the conditional probability, 
{\eqn{ \label{eq:1}
P_{\rm qu}(\xi |\xi_0)= P_{\rm qu}(s_4  | s_1,s_2,s_3) P_{\rm qu}(s_3 | s_1, s_2) 
= P_{\rm eq}(s_4  | s_2,s_3) P_{\rm eq}(s_3 | s_1, s_2),}}
On the r.h.s. of (\ref{eq:1}) we used the fact that the statistics of $s_4$ is independent of $s_1$ {\it if} $s_2$ and $s_3$ are specified.
We are going to show that
\eqn{  P_{\rm qu}(\xi |\xi_0)=P_{\rm eq}(\xi |\xi_0), \qquad  P_{\rm qu}(\xi )=P_{\rm eq}(\xi).  }
{\it Derivation ---}
Multiplying the both ends of (\ref{eq:1}) by $P_{\rm eq}(\xi_0),$
\beqa \label{eq:11}
P_{\rm qu}(\xi |\xi_0)P_{\rm eq}(\xi_0) 
&=&
P_{\rm eq}(s_4  | s_2,s_3) P_{\rm eq}(s_3 | s_1, s_2) P_{\rm eq}(\xi_0)
\cr &=&   P_{\rm eq}(s_4  | s_2,s_3) P_{\rm eq}(s_1,s_2,s_3) 
\cr &=&   P_{\rm eq}(s_4  | {s_1,}s_2,s_3) P_{\rm eq}(s_1,s_2,s_3) 
\cr &=&   P_{\rm eq}(\xi_0,\xi).
\eeqa
On the r.h.s. of (\ref{eq:11}) we used the fact that the statistics of $s_4$ is independent of $s_1$ if $s_2$ and $s_3$ are specified.
(\ref{eq:11}) means 
\eqn{ \label{eq:pcond21}
 P_{\rm qu}(\xi |\xi_0)=\frac{P_{\rm eq}(\xi_0,\xi)}{P_{\rm eq}(\xi_0)} =  P_{\rm eq}(\xi |\xi_0). 
}
Now admitting that the stationary probability $P_{\rm qu}(\xi)$ is the unique normalized solution of the Fredholm equation,
${ P_{\rm qu}(\xi )=\sum_{\xi_0}  P_{\rm qu}(\xi |\xi_0) P_{\rm qu}(\xi_0), }$
and that the equilibrium probability $P_{\rm eq}(\xi)$ is the unique normalized solution of 
${ P_{\rm eq}(\xi )=\sum_{\xi_0}  P_{\rm eq}(\xi |\xi_0) P_{\rm eq}(\xi_0), }$
the relation (\ref{eq:pcond21}) means that 
\eqn{P_{\rm qu}(\xi) =P_{\rm eq}(\xi)\,\,\,\,\, {\rm Q.E.D.}}

\subsection{Ising chain with up to the second-nearest neighbor interaction quenched two-after-two spins (Fig.\ref{fig:schema}(e))}
The first part of the argument is almost parallel as \S\ref{subsec:PQ11} except that the spin $s_i$ is replaced by the spin pair, $\xi_p$. 
Suppose that the {\sf Pq} is done by quenching the spin pair of the form, $\xi_p\equiv (s_{2p+1},s_{2p+2}).$ 

The {\sf Pq} in this model is a Markovian process for $\{\xi_p\}$ with $p$ playing the role of time. Following the argument 
of \S\ref{subsec:PQ11} line-to-line, we find that
\beq \label{eq:Pcond22} 
P_{\rm qu}(\xi_{p_0+1}|\xi_{p_0})= P_{\rm eq}(\xi_{p_0+1}|\xi_{p_0}),
\eeq
\beq \label{eq:Pcond22bis} 
P_{\rm qu}(\xi)=P_{\rm eq}(\xi),
\eeq
and, therefore, $P_{\rm qu}(\xi_{p_0},\xi_{p_0+1})= P_{\rm eq}(\xi_{p_0},\xi_{p_0+1}).$

We should still study separately the form of $P_{\rm qu}(s_{2p+2},s_{2p+3}|s_{2p},s_{2p+1})$
because the protocol of {\sf Pq} breaks the translational symmetry of the spin chain under the shift of a single spin position. 
 Again, we introduce the indexation of spin, $\ldots, s_1,s_2,s_3,s_4,s_5, \ldots = \ldots \xi_0, \xi,\ldots,$ where  $\xi_0=\{s_1,s_2\}$ and $\xi=\{s_3,s_4\}.$
In this notation, (\ref{eq:Pcond22}) and  (\ref{eq:Pcond22bis})
 means ${ P_{\rm qu}(s_1,s_2,s_3,s_4)=P_{\rm eq}(s_1,s_2,s_3,s_4).}$
The question is if ${ P_{\rm qu}(s_4,s_5|s_2,s_3)=P_{\rm eq}(s_4,s_5|s_2,s_3)}$ holds. The answer is yes. It suffices to show 
\eqn{P_{\rm qu}(s_2,s_3,s_4,s_5)=P_{\rm eq}(s_2,s_3,s_4,s_5).}
{\it Derivation --- } 
$P_{\rm qu}(s_2,s_3,s_4,s_5)=P_{\rm qu}(s_5|s_2,s_3,s_4)P_{\rm qu}(s_2,s_3,s_4)$ 
$=P_{\rm qu}(s_5|s_3,s_4)P_{\rm qu}(s_2,s_3,s_4)$$=P_{\rm qu}(s_5|s_3,s_4)$ $\times\sum_{s_1}P_{\rm qu}(s_1,s_2,s_3,s_4)$
$=P_{\rm eq}(s_5|s_3,s_4)$ $\sum_{s_1}P_{\rm eq}(s_1,s_2,s_3,s_4)$ $=P_{\rm eq}(s_5|s_3,s_4)$$P_{\rm eq}(s_2,s_3,s_4)$
$=P_{\rm eq}(s_5|s_2,s_3,s_4) $ $\times P_{\rm eq}(s_2,s_3,s_4),$
where, to go to the last equality, we used $P_{\rm eq}(s_5|s_3,s_4)$ $=$
$P_{\rm eq}(s_5|s_2,s_3,s_4)$ since in equilibrium
the statistics of $s_5$ is independent of $s_2$ if $(s_3,s_4)$ are specified. We, therefore, have
$P_{\rm qu}(s_2,s_3,s_4,s_5)=P_{\rm eq}(s_2,s_3,s_4,s_5).$ \,\,\,\,\, {\rm Q.E.D.}

\section{Conclusion}
What is in common between the present progressive quenching ({\sf Pq}) and the study in \cite{PQ-global-2017axv}  is the way we fixed the spins: We did it {\it as a snapshot} of the equilibrium state.
For the Ising chains with the interaction with the nearest neighbor or up to the second nearest neighbor spins, we found that {\sf Pq} of a single spin or a pair of neighboring spins generates the ensemble of spin configurations which is identical to the equilibrium ensemble of the given system. 
It is somehow counter-intuitive that the non-equilibrium and inhomogeneous operation of {\sf Pq} leaves the equilibrium ensemble intact. The evident source of the equilibrium ensemble is that, in our protocol, the unquenched part of the system is equilibrated before the quenching of spin or spins and, moreover, the {\it spatial} Markovian nature of the equilibrium fluctuations should be essential.
In the case of globally coupled Ising model \cite{PQ-global-2017axv}, the ensemble generated by the {\sf Pq} is qualitatively different from the equilibrium one. In the latter case, however, there emerged a quasi-martingale property in the unquenched equilibrium spin, $m_T$ in their notation, which reflect the fact that the quenching is done as a snapshot of the equilibrium state.
 
As  non-trivial extensions of the present study, we may adapt a kinetic Ising model either of Glauber \cite{GlauberDynamics} or of Kawasaki \cite{kawasaki-bracket}. Then a characteristic length should intervene in the quenched ensemble, as it was the case for the phason system \cite{phason-freezing-PhA}.

\section*{Acknowledgement}
KS thanks the organizers of the the 30th Marian Smoluchowski Symposium (September, 2017, Krakow).
We thank the laboratory Gulliver, ESPCI for welcoming the training course during which this work has been accomplished.

\bibliographystyle{apsrev4-1.bst}   
\bibliography{PQ-2nnIsing-bib}
\end{document}